\documentclass[twocolumn,twoside,slac]{revtex4}
\usepackage{graphicx}
\usepackage{fancyhdr}
\begin{document}

\title{CDF Detector Simulation Framework and Performance}

%

\author{E. Gerchtein, M. Paulini}
\affiliation{Carnegie Mellon University, Pittsburgh, PA 15213, USA}

\begin{abstract}
                                           
The CDF detector
simulation framework is integrated into an AC++ 
application used to process events in the CDF
experiment. The simulation framework is based on the GEANT3 package. 
It holds the detector element geometry
descriptions, allows configuration of digitizers at run-time and manages 
the generated data. The design is based on
generic programming which allows for easy extension of the simulation
framework.  
The overall design, details of
specific detector components and in particular the performance 
of the CDF simulation compared to collider data
are described.

\end{abstract}

\maketitle

\thispagestyle{fancy}


\section{Introduction}

The Collider Detector at Fermilab
(CDF)~\cite{cdf} operating at the Run\,II Tevatron Collider is
a complex general purpose detector and has many subsystems. 
For the simulation of each subdetector, one needs geometry, a digitizer to
generate hits, 
a menu to change digitizer's configuration, and an object to keep a simulated
event. The simulation framework allows to plug these pieces into the 
existing software system and hides complicated details of their interaction
from the 
user.
To achieve these goals, the framework design is based upon a
mixture of generic programming and object oriented principles.
This mix, rather than a pure object oriented solution, allows
the framework to be easily expandable while retaining as much time 
efficiency as possible.

\section{General overview}

The CDF detector simulation framework is integrated into an AC++~\cite{ac++}
application used to process events at the CDF experiment. The tracking of
particles  
through matter is performed by the GEANT3~\cite{geant3} package.
The CDF software uses the same geometry 
for event reconstruction and simulation. The CDF geometry package provides 
the volume description and geometry tree creation. 
The CDF geometry was  designed to allow seamless 
transition to GEANT4 if desired. It can be converted to a GEANT3
or GEANT4 geometry. An output event of the simulation contains data in the 
same format as raw detector data plus Monte Carlo truth information.

\subsection{Functionality of cdfSim executable}

The main simulation executable, cdfSim, allows generation of physics events
with different generators such as Herwig v6.5, PYTHIA v6.2, Isajet v7.51,
WGRAD, WBBGEN, GRAPPA (GRACE for $p \bar{p}$), Vecbos, BGenerator, 
MinBiasGenerator, SingleParticle. Les Houches Accords - a universal
interface between matrix element generators and parton shower MC programs - is
implemented in Herwig, PYTHIA and GRAPPA. 

In addition to decay routines internal to each MC generator, particle
decays can be simulated by three 
decay packages - QQ v9.1, EvtGen and Tauola.

The cdfSim executable may be used with different configurations of
subdetectors, 
different geometry levels, and physics processes  depending on
desired accuracy versus time efficiency. The following sub-detectors
are simulated:
\begin{itemize}
\item Silicon detectors (SVX, ISL)
\item Central Outer Tracker (COT)
\item Muon systems
\item Time-of-Flight system (ToF)
\item Calorimeters
\item Cherenkov Luminosity Counters (CLC)
\item Forward detectors (Miniplug, BSC, RPS) 
\end{itemize}
In addition, the CDF geometry contains a detailed description of passive
material elements, in particular within the silicon detectors.

\subsection{ Generators within AC++ framework }
 
\begin{figure*}[tb]
\centering
\includegraphics[width=105mm]{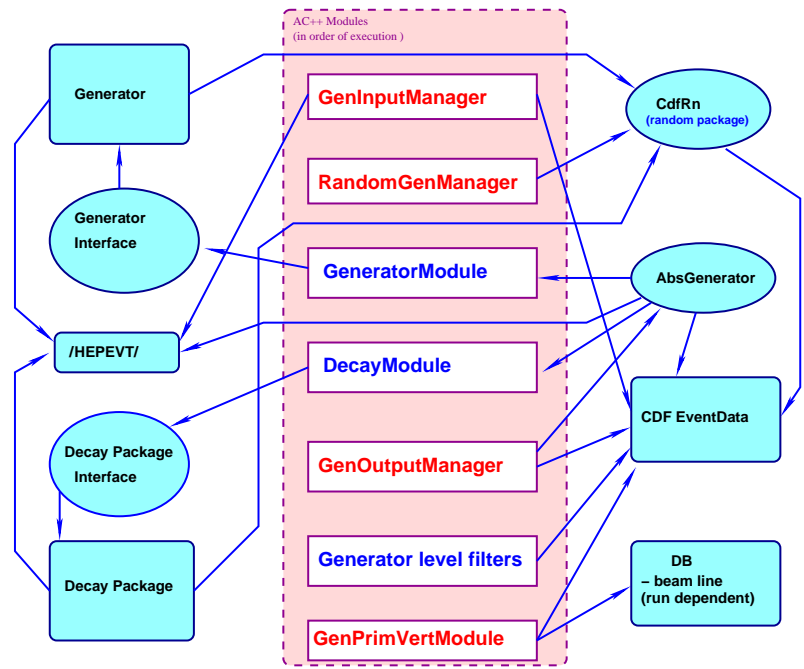}
\caption{Generator related modules within the AC++ framework.} 
\label{sim-f1}
\end{figure*}

The flow diagram in Figure~\ref{sim-f1} shows 
how the various generators are incorporated
within the AC++ framework. The AC++ modules are shown in order of execution
from top to bottom. A HEPEVT common block is used for communication between 
different generators and decay packages. The GenInputManager module 
creates an instance of 
the CdfRn class which interfaces to the CLHEP random number generators. 
CDF unified the usage of the various
random number generators used throughout the simulation code including all
generator packages.
This ensures statistically independent production of
large Monte Carlo samples on the CDF Production Farms. 
CdfRn manages the random engines of its clients and allows 
restoring and saving of 
current engine configurations. 
The GenInputManager and RandomGenManager are followed by a sequence
of generator modules and decay modules. 
The combination of enabled modules from this sequence 
is checked at the  AC++ framework level for user setup errors.
Each module has an interface to the 
generator/decay package controlling the access to the underlying FORTRAN
routines.
The generator or decay package modifies the HEPEVT common block. 
After the event has been processed
by all enabled generators and decay packages, the
HEPEVT common block is converted into a persistent object and  
added to the event record. After this step, a sequence of generator level
filters can be evoked to
reject events based on user chosen selection criteria. 
Finally, the position of the primary interaction vertex is generated by the
GenPrimVert module which allows to
take beam line information from the CDF data base depending on a chosen run
number. 
This information is also added to the event record.

\subsection{ Simulation framework }

\begin{figure*}[tb]
\centering
\includegraphics[width=110mm]{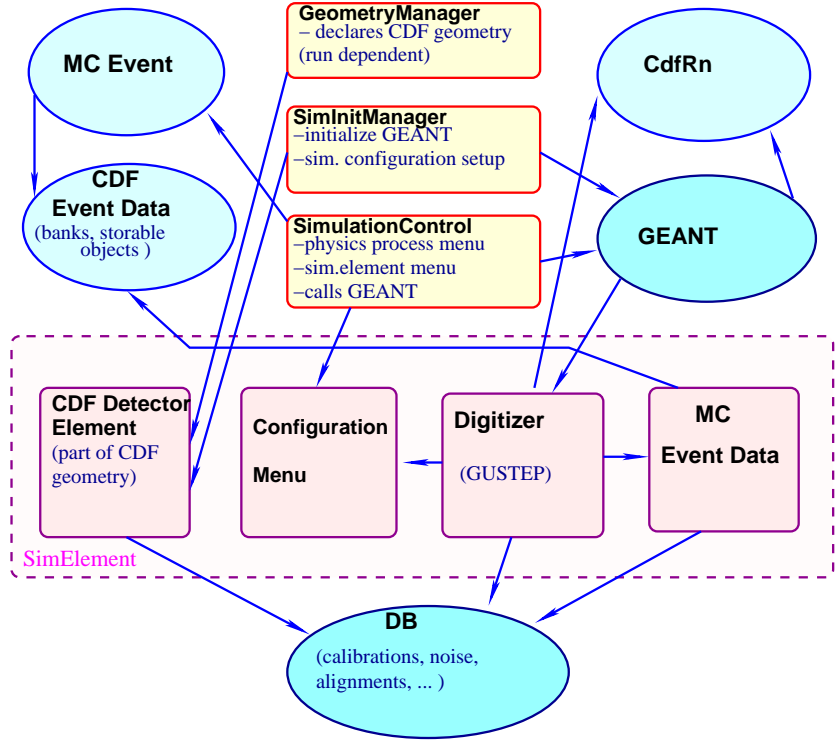}
\caption{Generator related modules within the AC++ framework.} 
\label{sim-f2}
\end{figure*}

The diagram in Figure~\ref{sim-f2} shows the relationship between different
components 
of the simulation framework. The boxes represent AC++ modules in order of
execution 
from top to bottom. An instance of the CDF geometry is created by 
the GeometryManager
module at the beginning of the job. The SimInitManager module initializes 
GEANT3, passes
the CDF geometry to GEANT3, and obtains the information which simulation
configuration is used. The SimulationControl module instantiates 
the simulation elements
requested by the SimInitManager. The simulation elements are processed by
a simulation base which is implemented as an abstract factory owned by
SimulationControl. Each simulation element consists of a geometry,
a configuration menu, a digitizer, and an object to keep the Monte Carlo
event data. 

During the event processing the SimulationControl module interacts with 
GEANT3 and passes to it the Monte Carlo particles from the generator. 
Inside an  active detector volume, GEANT3 calls 
a user defined stepping routine and control is passed to the digitizer of the
corresponding simulation element. This action is dispatched by the
simulation base. 
The digitizer creates hits, depending on the configuration menu, 
and adds the information 
to the Monte Carlo event data. 
After tracing of particles with GEANT3 is finished, the hit information 
from the MC event data object is converted into the persistent object and 
added to the event record. 

At the beginning of each simulation run, information from the CDF data base
may be accessed to obtain noise,  
calibrations or alignment constants used for the simulation job.  

\section{ Simulation performance }

In the following subsections the current status of the simulation of the
different subdetectors 
is described including a discussion of the simulation performance compared
to collision data. 

\subsection{ Silicon detector }

The Run\,II silicon 
detector consists of seven double sided layers and one single sided layer
mounted on the beampipe covering a total radial area from 1.5-28~cm. The
silicon vertex detector covers the full Tevatron luminous 
region and allows
for standalone silicon tracking up to a pseudo-rapidity $|\eta|\sim2$.
The geometry of the silicon detectors has been implemented in great details
in the simulation including passive materials and allows for detector
misalignment. 

The main purpose of simulating the silicon detector is to describe the
charge deposition of traversing tracks
on the silicon strips well enough to evaluate silicon tracking
performance. The basic distributions to model are the number of strips
included in a charge cluster and the cluster shape.

Three charge deposition models (CDM) are implemented in the simulation
framework: geometric, parametric and physical. 
The geometric model is based purely on geometry. 
The total amount of charge deposited on a sensor is calculated by GEANT3 based
on an unrestricted Landau distribution.
The amount of charge deposited on each strip is a fraction of the entire
charge  
proportional to the path length of the particle in the vicinity of the strip.
The result of this method is that for tracks at a certain incident angle,
all strips in the cluster have the same amount of charge 
deposited. 
The cluster length and shape in silicon data is considerably
more complicated due to various physics effects such as $\delta$-rays.

The parametric model is based on the physics effects simulated by GEANT3.
However, the GEANT3 results have been parameterized into template
distributions to increase the speed
of the code. The important effects that are included in this model are
energy deposition using a restricted Landau distribution with $\delta$-ray
production above a cutoff value, capacitive charge sharing between strips,
and noise. This CDM has been further tuned to data.

The physical model simulates the physics of charge deposition from first
principals. The basic idea is to have the capability of monitoring, 
predicting and comparing the electrical behavior of the real detector
with the expectations from theory. The physics described by this model 
includes continuous energy loss using a restricted Landau distribution with
$\delta$-ray production, convolution of Landau fluctuations with a Gaussian,
magnetic field effects, diffusion of electrons, holes, and noise. 
In addition to these effects, the charge inversion of a sensor can be
simulated. 

\begin{figure}[tbh]
\centering
\includegraphics[width=80mm]{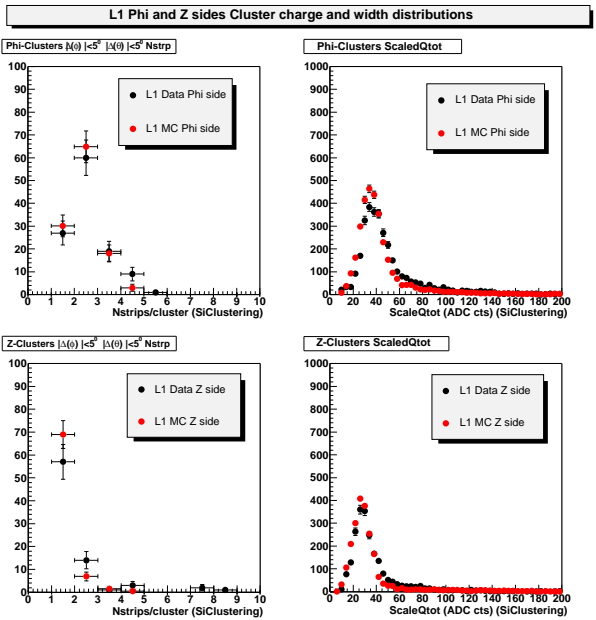}
\caption{Cluster length (left) and cluster profile (right) for the
axial side (top) and stereo side (bottom) of Layer\,1 of SVX\,II.
Data are the black points, Monte Carlo are the shaded (red) points. } 
\label{svx-f1}
\end{figure}

\begin{figure}[tbh]
\centering
\includegraphics[width=80mm]{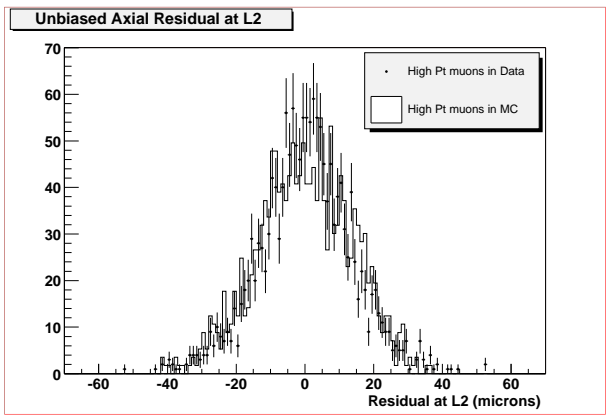}
\caption{Unbiased axial residuals (Layer\,2). 
Date are the points with error bars; MC is the histogram. } 
\label{svx-f2}
\end{figure}

In Figure~\ref{svx-f1} CDF 
silicon data are compared with simulation using the parametric CDM.
The cluster length (left) and cluster profile (right) are compared for both
the axial side (top) and stereo side (bottom) of Layer\,1.
The intrinsic silicon resolution 
is displayed in Figure~\ref{svx-f1}.
Unbiased axial residuals (Layer\,2) are compared between
data and MC.
Silicon simulation and data are in good agreement.

\subsection{Central Outer Tracker}

CDF's Central Outer Tracker contains 30,200 sense wires
arranged in 96 layers combined into four axial and four stereo
superlayers. It also provides d$E$/d$x$ information for particle
identification. 
The default drift model used in the COT simulation is based on the GARFIELD
package~\cite{garfield}, 
a general drift chamber simulation program. 
The default GARFIELD parameters are scaled to describe the data.

\begin{figure}[tb]
\centering
\includegraphics[width=80mm]{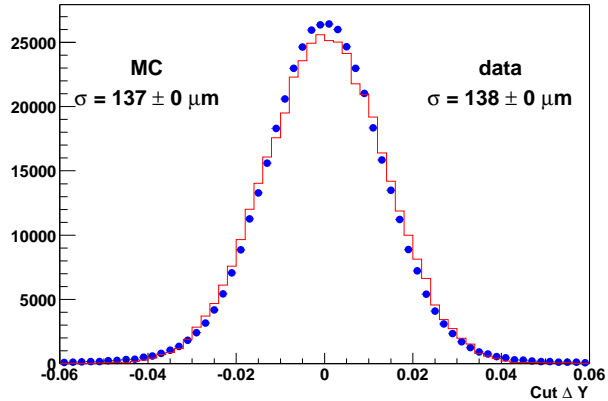}
\caption{Track residuals (hit displacement minus track
  displacement) for data (points) and MC (histogram).}  
\label{cot-f1}
\end{figure}

An example of the performance of COT simulation is displayed in 
Figure~\ref{cot-f1} 
which shows a comparison of track residuals between data
and MC. The track residual is calculated as the hit displacement 
minus track displacement after applying track quality cuts. 
The simulation (histogram)
is in a good agreement with the data (points). 

\begin{figure}[tb]
\centering
\includegraphics[width=80mm]{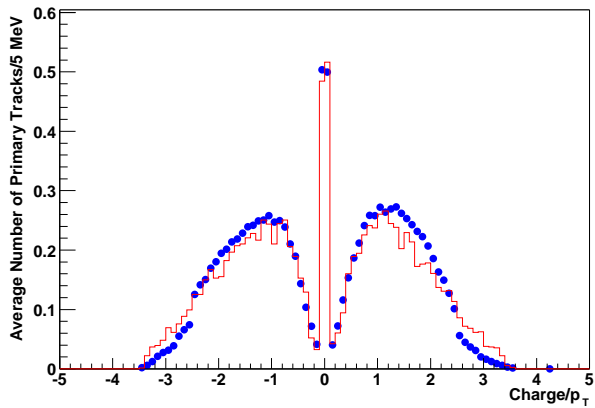}
\caption{Signed track curvature (charge/$p_T$) for  
primary tracks with impact parameter $d_0<3$~mm in $W\rightarrow \mu\nu$
data (points) and MC events (histogram).}  
\label{cot-f2}
\end{figure}

Figure~\ref{cot-f2} is a comparison of $W \rightarrow \mu \nu$ events 
generated with PYTHIA to $W$~data obtained from an inclusive 
high $p_T$ muon data set with a muon trigger threshold of 18~GeV/$c$.
Shown is the distribution of signed track curvature (charge/$p_T$) for  
primary tracks with impact parameter $d_0<3$~mm.
The data (points) are reasonably well described by the
Monte Carlo (histogram). 

\subsection{ Time-of-Flight system }

\begin{figure}[tb]
\centering
\includegraphics[width=80mm]{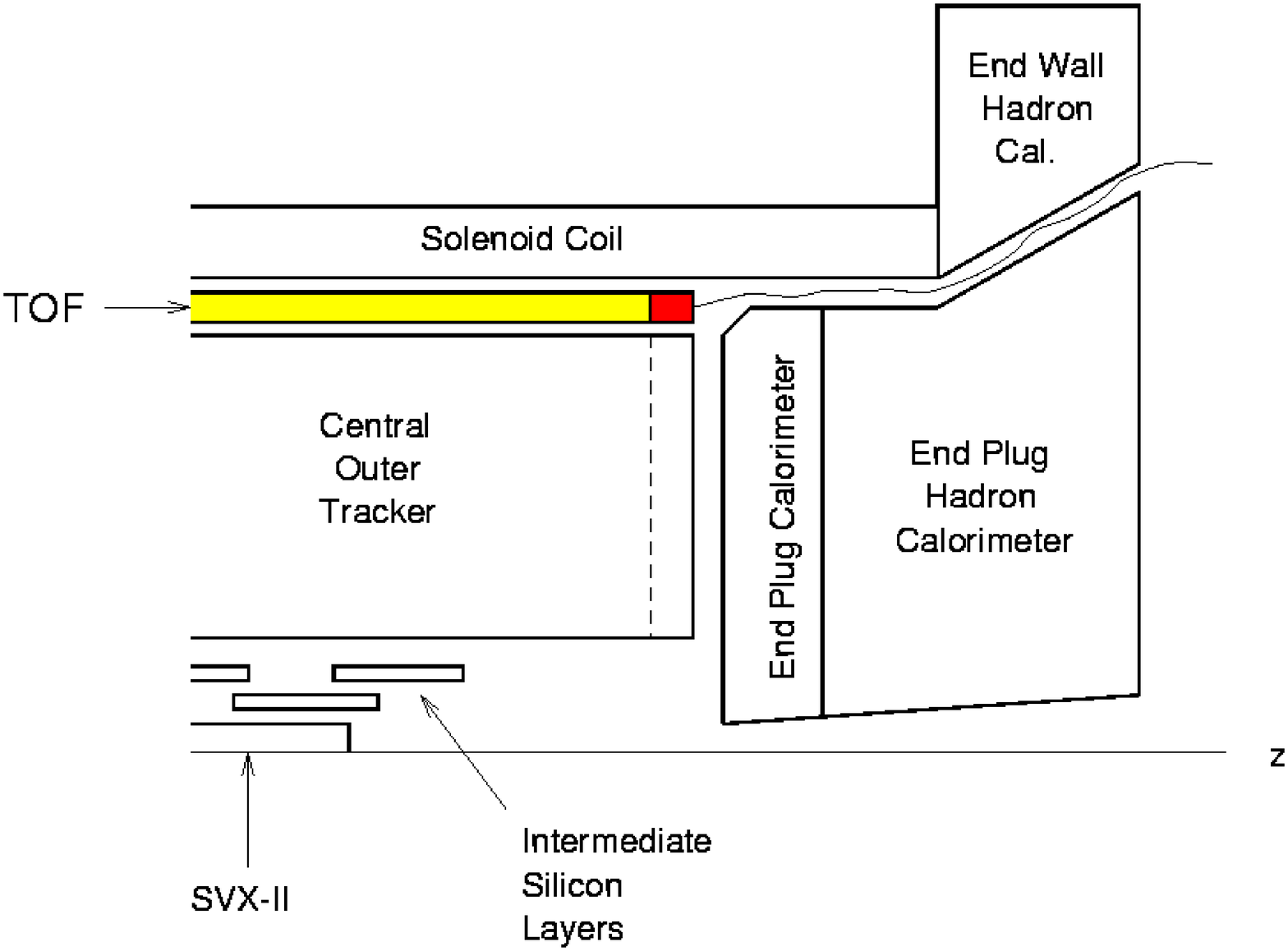}
\caption{Schematic view of the CDF detector indicating the location 
of the scintillator bars and attached PMT's.
} 
\label{tof-f1}
\end{figure}

One of the new devices in the CDF\,II detector is a
Time-of-Flight system with a resolution of about
100~ps. It employs 216 three-meter-long
scintillator bars with photomultiplier tubes (PMT's) at each end located
between the outer radius of the COT 
and the superconducting solenoid (see Figure~\ref{tof-f1}).
The Time-of-Flight system is designed for the identification
of kaons with a 2\,$\sigma$-separation between $\pi$ and $K$ for
$p<1.6$~GeV/$c$.

The energy deposition of the incident 
particle as well as the time at entry and exit points are simulated by
GEANT3. 
This  information is translated into 
TDC and ADC counts. Three response parameterizations with different levels
of detail in the 
parameterization are implemented. 
One of the models uses 
a simple analytic form to parametrize time and ADC count. 
Two other models use calibration parameters from ToF data.

\begin{figure}[t]
\centering
\includegraphics[width=80mm]{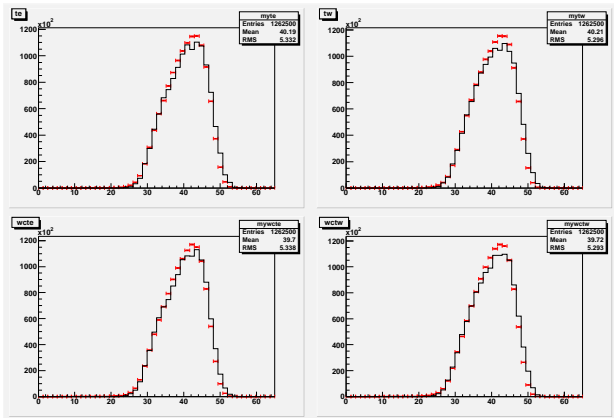}
\caption{Comparison of time distribution between 
the data and simulation. Crosses with the error bars 
are data; the histogram is simulation. The upper 
(lower) plots are a comparison of the time distribution for the east 
(west) channels before (top)
and after (bottom) time-walk correction. 
} 
\label{tof-f2}
\end{figure}

Figure~\ref{tof-f2} shows a comparison of the time distribution between 
collision data and simulated events. Crosses with error bars 
denote data and the histogram represents the simulation. The upper 
(lower) plots are a comparison of the time distribution for the east 
(west) channels before (top)
and after (bottom) time-walk correction. 
The comparison indicates reasonable agreement between data and simulation
allowing for further tuning of the simulation.


\subsection{ Muon systems }

\begin{figure}[t]
\centering
\includegraphics[width=80mm]{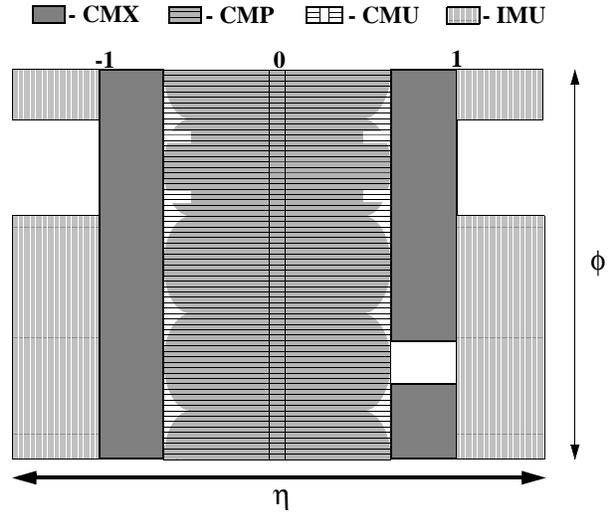}
\caption{Location of the CDF muon systems in azimuth $\phi$
and pseudorapidity $\eta$. } 
\label{muon-f1}
\end{figure}

There are four muon sub-systems in CDF: Central Muon Detector (CMU),
Central Muon Upgrade detector (CMP), Central Muon Extension (CMX),
and Intermediate Muon Detector (IMU). 
Figure~\ref{muon-f1} shows the 
location of the muon systems in azimuth $\phi$ and
pseudorapidity $\eta$. 

The central calorimeters act as a hadron absorber for the muon
detection system. 
The CMU consisting of four layers of planar drift 
chambers is located directly outside the central calorimeters.
The CMU system covers $|\eta| \leq 0.6$ and can be reached by muons
with $p_T$ in excess of 1.4~GeV/$c$.
To reduce the probability of misidentifying penetrating hadrons as muon
candidates in the central detector region,
four additional layers of drift chambers (CMP) 
are located 
behind 0.6~m of steel outside the CMU system. To reach these two detectors,
particles produced at the primary interaction vertex must traverse material
totaling 5.4 and 8.4 pion interaction lengths, respectively. 
The CMX is located in the pseudorapidity
interval $0.6 < |\eta| < 1.0$ extending the polar acceptance of the
muon system. 
This detector consists of free-standing conical 
sections at the four corners of the central detector with eight 15 degree
wedges. 
The IMU identifies muons over the rapidity range $1.0<|\eta|<1.5$ by a 4-layer
barrel of drift tubes (BMU) and scintillators (WSU, BSU, TSU).
All muon systems have ADC/TDC read-out.
The Run\,II upgrades to the muon system almost double the central
muon coverage compared to Run\,I and extent it up to $|\eta|\sim1.5$.

\begin{figure}[tb]
\centering
\includegraphics[width=80mm]{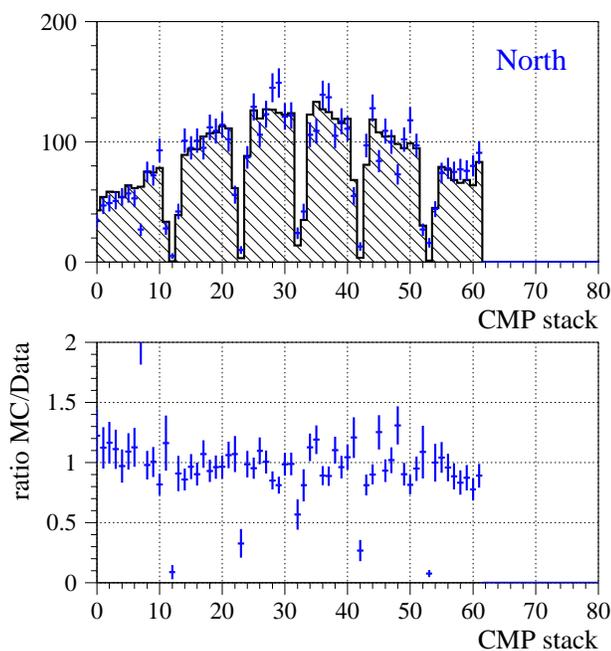}
\caption{Top: Comparison of the number of CMUP muons for each stack
in the North wall, for data (points with error bars) and the simulation (shaded
histogram). Bottom: Ratio of the above distributions.
} 
\label{muon-f2}
\end{figure}

The most challenging part in the muon simulation is the description of the
complicated geometry of the muon systems, 
while tuning of the digitization to muon trigger data is straight forward. 
Special efforts were made to compare simulation to real data 
with the aim of checking the geometry description. 
The study was done for CMP, CMU, and CMX using muons from
$W \rightarrow \mu \nu$ and $Z   \rightarrow \mu^+ \mu^-$ in data and
simulation. 
The overall description of the data is good. 
As an example of the numerous studies, 
Figure~\ref{muon-f2} shows a comparison of simulation and data
for the number of muons per stack in the North wall.  

\subsection{Cherenkov Luminosity Counters}

\begin{figure}[tb]
\centering
\includegraphics[width=80mm]{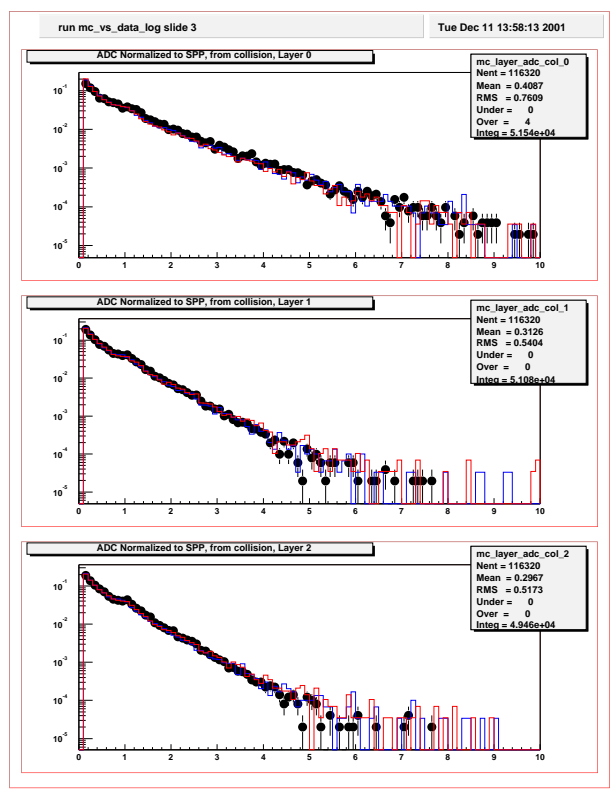}
\caption{Amplitude distributions in the three layers of CLC counters. 
Dots are Monte Carlo,
the red (west) and blue (east) histograms are data.
}
\label{clc-f1}
\end{figure}

The Cherenkov Luminosity Counters~\cite{clc} are used to measure the
Tevatron luminosity at CDF. 
The CLC acceptance for $p \bar{p}$ inelastic processes is estimated from the
simulation and adds a major contribution  
to the luminosity uncertainty. 
An excellent CLC simulation performance is therefore critical.
The generation and propagation of Cherenkov photons is fully simulated by
GEANT3 and then corrected for photo-cathode efficiencies.
The hit count in the CLC is sensitive to the material traversed by particles
between 
the interaction point and the CLC counters. 
Therefore, the detector geometry in front of and in the vicinity
(back-scattering) of the CLC needs to be   
described with high accuracy. 
Monte Carlo and data are in good agreement as can be seen in 
Figure~\ref{clc-f1}. It shows a comparison of amplitude distributions 
in the three layers of the CLC counters. Dots are Monte Carlo (PYTHIA), 
the red (west) and blue (east) histograms 
are minimum bias data. 
The distributions are normalized to the single particle peak. 
The resulting uncertainty on the CDF luminosity calculation due to the CLC
simulation 
is less than $4\%$. 

\subsection{ Calorimeters }

\begin{figure}[tb]
\centering
\includegraphics[width=65mm]{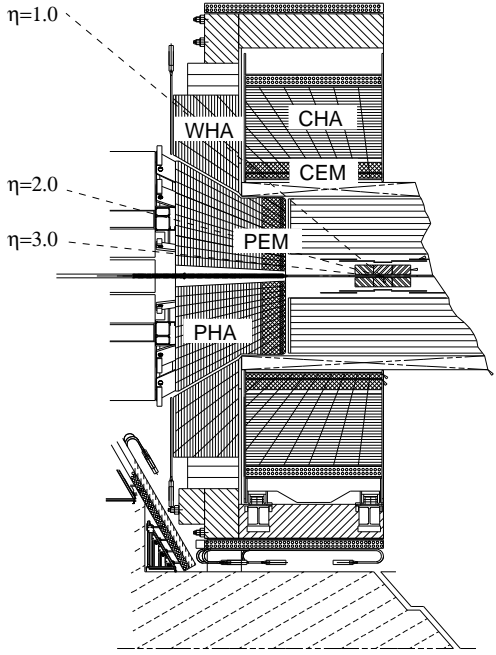}
\caption{Elevation view of one half of the CDF detector displaying 
the various calorimeter compartments.
} 
\label{cal-f1}
\end{figure}

The CDF calorimeters~\cite{cal1,cal2,cal3} provide 
separate electromagnetic (EM)
and hadronic (HAD) measurements as shown in Figure~\ref{cal-f1}. 
The central electromagnetic (CEM) and hadronic (CHA)
calorimeters ($|\eta|<1.1$) employ a projective tower geometry
back to the nominal interaction point with a 
segmentation of $\Delta\eta \times \Delta\phi \sim 0.1 \times 15^{\circ}$.
The sampling medium is composed of scintillators layered with lead
absorbers in the electromagnetic calorimeter and steel in the CHA. 
The energy resolution for the CDF
central calorimeter is 
$\sigma(E_T) / E_T = [(13.5\% / \sqrt{E_T})^2 + (2\%)^2]^{1/2}$ 
for electromagnetic showers and
$\sigma(E_T) / E_T = [(50\% / \sqrt{E_T})^2 + (3\%)^2]^{1/2}$ 
for hadrons with $E_T$ measured in GeV.
A layer of proportional chambers (CES), with wire and strip readout, 
is located six radiation lengths deep in the CEM calorimeter,
approximately near shower~maximum for electromagnetic showers.
The CES provides a measurement of electromagnetic shower profiles 
in both the $\phi$- and $z$-directions.
Proportional chambers located between the solenoid and the CEM
comprise the central preradiator detector (CPR), which samples the
early development of electromagnetic showers in the material of the
solenoid coil, providing information in $r$-$\phi$ only.
In addition, the
forward calorimeters have been replaced in Run\,II by a new scintillator
tile based plug calorimeter which covers
the range $|\eta| < 3.6$ and
gives good electron identification up to $|\eta| \sim 2$.
The plug 
electromagnetic calorimeter also has fine grained shower profile detectors
at electron shower maximum, and preshower pulse height detectors
at approximately $1 X_o$ depth. 
The calorimeter cell segmentation is summarized in
Table~\ref{calseg}.  A comparison of the central and plug
calorimeters is given in Table~\ref{plugcomp}.
\begin{table}[tb]
\begin{center}
\begin {tabular} {l|c|c}
\hline 
$|\eta|$ Range & $\Delta\phi$ & $\Delta\eta$ \\
\hline
0. - 1.1 (1.2 h) & $15^{\circ}$ &  $\sim0.1$ \\
1.1 (1.2 h) - 1.8 & $7.5^{\circ}$ & $\sim0.1$ \\
1.8 - 2.1 & $7.5^{\circ}$ & $\sim0.16$ \\
2.1 - 3.64 & $15^{\circ}$ & 0.2 $-$ 0.6 \\
\hline 
\end{tabular}
\caption{CDF Run\,II calorimeter segmentation.}
\label{calseg}
\end{center}
\end{table}
\begin{table}[tb]
\begin{center}
\begin {tabular} {l|c|c}
\hline 
 & Central  & Plug \\
\hline
EM: & &\\
Thickness & $19 X_0, 1 \lambda$ & $21 X_0, 1 \lambda$ \\
Sample (Pb) & $0.6 X_0$ & $0.8 X_0$  \\
Sample (scint.) & 5 mm & 4.5 mm \\
WLS & sheet & fiber \\
Light yield & $160$ pe/GeV & $300$ pe/GeV \\
Sampling res. & $11.6\%/\sqrt{E_T}$ & $14\%/\sqrt{E}$ \\
Stoch. res. & $14\%/\sqrt{E_T}$ & $16\%/\sqrt{E}$ \\
Shower Max. seg. (cm)& 1.4$ \phi \times $(1.6-2.0) Z& $0.5\times0.5$ UV\\
Pre-shower seg. (cm)& $1.4 \phi\times 65$ Z & by tower \\
\hline
Hadron: &&\\
Thickness& $4.5 \lambda$  & $7 \lambda$\\
Sample (Fe)&1 to 2 in.& 2 in.\\
Sample (scint.)& 10 mm & 6 mm\\
WLS & finger & fiber\\
Light yield& $\sim40$ pe/GeV  & $39$ pe/GeV\\
\hline 
\end{tabular}
\caption{Comparison of CDF Run\,II central and plug calorimeters.
}
\label{plugcomp}
\end{center}
\end{table}

The main objective for the calorimeter 
simulation is speed.
The simulation of the CDF calorimeters is based on the GFLASH~\cite{gflash} 
parameterization package interfaced with the GEANT3 simulation.
The simulation of 
EM and HAD showers in GFLASH is  initiated when particles undergo inelastic 
interactions inside the calorimeter volume.  GFLASH uses a mixture-level 
GEANT3 geometry description for the CDF calorimeters. 
It handles the spatial distribution
of the deposited energy and
the energy loss of particles inside
the geometry volumes.

The GFLASH parameters are tuned.
The basic idea of the tuning process is first to bring
the response of the electromagnetic calorimeter into agreement with data and
then tune the hadron response by adjusting GFLASH parameters to match pion
testbeam data and $p\bar p$~collision data.
The tuning of GFLASH is split according to the two sets of parameters 
that control the fraction of visible energy produced in the active medium,
and the energy and depth dependent spatial distribution 
of the various components of the shower model.

To tune EM showers, 8-227~GeV electrons from test beam data were used.
The energy scale and resolution for electrons was verified using $E/p$.
The hadronic shower shape was tuned to test beam data and minimum bias data.
High $p_T$ tuning was done using high  $p_T$ pions test beam data 
in the energy range 7-227~GeV (mainly 57~GeV) for CHA
and 8-227~GeV for PHA. The low $p_T$ tuning process uses 
isolated low $p_T$ tracks from minimum bias data.
The Minimum-Ionizing-Particle (MIP) peak is measured in the EM compartments
with pions. 
The hadronic energy scale in the tuning procedure is set by the 
response of 57~GeV pions in the central and plug calorimeters.
Since there are no test beam data available for the wall calorimeter, 
the energy scale of WHA is set to CHA. Once the MIP distribution 
and the hadron energy scale are set, the logarithmic energy dependence of the
calorimeter response can be tuned.
Finally, for the tuning of the lateral shower profile, tracks obtained from
minimum bias events 
measured in the central part of the CDF detector are used. The available 
energy range was 0.5-2.5~GeV.


\begin{figure}[tb]
\flushleft
\includegraphics[width=80mm]{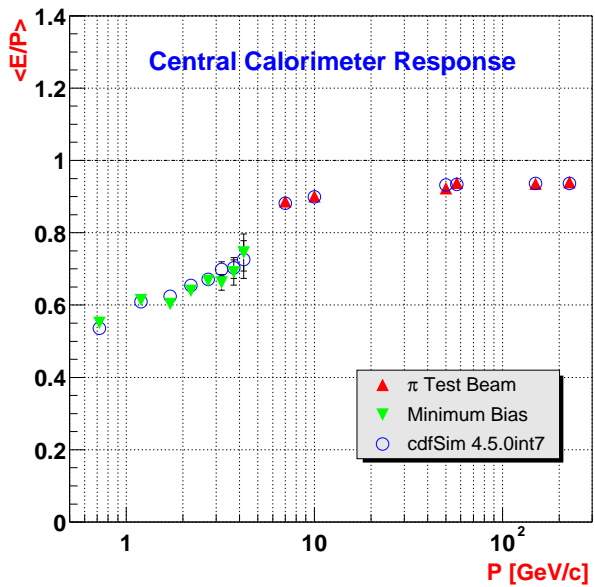}
\caption{ 
Comparison of $E/p$ distributions between simulation and testbeam as well as
collision data for different particle momenta.
} 
\label{cal-f2}
\end{figure}
\begin{figure}[t]
\flushleft
\includegraphics[width=80mm]{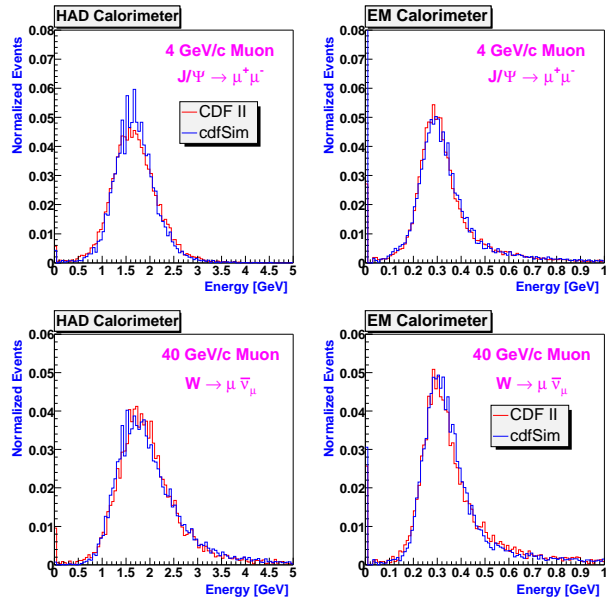}
\caption{ 
Comparison of energy depositions in the CDF hadronic (left)
and electromagnetic (right) calorimeter for 4~GeV$/c$ (top) and 40~GeV$/c$
(bottom) muons.
}
\label{cal-f3}
\end{figure}

Figure~\ref{cal-f3} 
Figure~\ref{cal-f2} shows a comparison of 
the $E/p$ distributions between simulation and testbeam as well as
collision data at low $p$ for different particle momenta.
The calorimeter response in the simulation yields excellent agreement with
data. 
As another example of calorimeter simulation performance, 
compares the energy deposition
of muons in the CDF calorimeter.  Data are shown as dashed lines and MC as
solid. The two top plots show
the energy distributions for 4~GeV$/c$ muons from 
$J/\psi\rightarrow\mu^+\mu^-$ in the 
hadronic (left) and electromagnetic (right) calorimeters.  The two bottom
plots  show
these distributions for 40~GeV$/c$ muons from
$W\rightarrow\mu\nu$~decays. Again, 
excellent agreement is seen between data and Monte Carlo.

The GFLASH based calorimeter simulation is about one hundred times faster
then the GEANT3 shower simulation for a typical top quark event.

 
\section{Conclusion}

We described the overall design of the CDF detector
simulation framework which is based on the GEANT3 package and integrated
into an AC++ application used to process events in the CDF
experiment. 
The CDF simulation framework is shown to be flexible, easily extensible,
and efficient. It is hiding complex infrastructure details from a user and
suitable for large scale Monte Carlo production. 
Subdetectors are successfully implemented
within the CDF simulation framework. 
We discussed details of the simulation of
specific detector components and in particular the performance 
of the CDF simulation which shows good agreement with 
$p\bar p$~collider data. 
A further upgrade to GEANT4 is possible through the low level API 
without affecting the client's code.

\begin{acknowledgments}

The CDF detector simulation project is the summary of a collective effort
of CDF institutions working 
on implementing and tuning simulation for the many CDF subdetectors. 
The authors would like to thank all the members of the CDF simulation 
group for their support and fruitful discussion.
This work is supported by the U.S.~Department of Energy under grant
DE-FG02-91ER40682. 

\end{acknowledgments}


\end{document}